Tal I. Sommer and Ori Katz

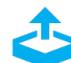

View Online    Export Citation    CrossMark

### ARTICLES YOU MAY BE INTERESTED IN

Crystal-like topological defect arrays in nematic liquid crystal
Applied Physics Letters **119**, 121906 (2021); https://doi.org/10.1063/5.0064303

Polarization switching in $Hf_{0.5}Zr_{0.5}O_2$-dielectric stack: The role of dielectric layer thickness
Applied Physics Letters **119**, 122903 (2021); https://doi.org/10.1063/5.0056448

Coarsening of polycrystalline patterns in atomically thin surface crystals
Applied Physics Letters **119**, 123102 (2021); https://doi.org/10.1063/5.0055078

# Pixel-reassignment in ultrasound imaging

Tal I. Sommer[1,2]  and Ori Katz[1,a)]

### AFFILIATIONS

[1]Department of Applied Physics, Hebrew University of Jerusalem, Jerusalem 9190401, Israel
[2]Alexander Grass Center for Bioengineering, Hebrew University of Jerusalem, Jerusalem 9190401, Israel

[a)]Author to whom correspondence should be addressed: orik@mail.huji.ac.il

### ABSTRACT

We present an adaptation of the pixel-reassignment technique from confocal fluorescent microscopy to coherent ultrasound imaging. The method, ultrasound pixel-reassignment (UPR), provides a resolution and signal to noise (SNR) improvement in ultrasound imaging by computationally reassigning off-focus signals acquired using traditional plane wave compounding ultrasonography. We theoretically analyze the analogy between the optical and ultrasound implementations of pixel reassignment and experimentally evaluate the imaging quality on tissue-mimicking acoustic phantoms. We demonstrate that UPR provides a 25% resolution improvement and a 3 dB SNR improvement in *in vitro* scans without any change in hardware or acquisition schemes.

Published under an exclusive license by AIP Publishing. https://doi.org/10.1063/5.0062716

Deep tissue high-resolution imaging is an indispensable tool for early detection and diagnosis of various medical conditions, as well as for basic biomedical research. One of the leading noninvasive approaches for imaging tissues is ultrasound echographic imaging, as ultrasound waves can noninvasively penetrate to large depths. However, the resolution of ultrasound imaging is fundamentally limited by the diffraction-limit, which is set by the ultrasound frequency and the imaging geometry. While the highest imaging resolution would be obtained for the highest ultrasound frequency, increasing the ultrasound center frequency results in a lower penetration depth, since the ultrasound attenuation coefficient in biological tissue grows with the ultrasound center frequency.[1] As a result, imaging at large depths suffers from a lower signal-to-noise ratio (SNR) and resolution.[2]

Various approaches for improving the resolution of ultrasound imaging have been put forward over the years. These include computational approaches such as deconvolution[3] and alternative reconstruction algorithms,[4–7] which usually improve resolution at the price of a reduction in SNR; structured illumination,[8–10] which requires a specific sonication pattern for each target depth; and recent approaches that correct for variations in speed-of-sound.[11–16] However, most of these approaches effectively utilize confocal-detection, which, as in optical microscopy,[17,18] improves resolution and background rejection, at the price of reduced signal collection efficiency.

A recently introduced method termed image-scanning microscopy[19,20] (ISM) or pixel-reassignment[21] allows full use of all signal photons without sacrificing transverse imaging resolution. Moreover, depending on the imaging point spread function (PSF), ISM can also provide an improvement in imaging resolution.[22,23]

Here, we adapt ISM to ultrasound imaging (Fig. 1) and demonstrate that it allows simultaneous improvements in both SNR and resolution. Importantly, while the optical implementation of ISM requires additional detectors compared to confocal microscopy, the ultrasound implementation of ISM does not require any change in hardware or acquisition schemes. The improvement in resolution and SNR is obtained by processing conventional plane wave compounding[24] ultrasound imaging data. We term this technique ultrasound pixel-reassignment (UPR).

In conventional confocal imaging, a focused illumination beam scans the field-of-view (FOV) point-by-point. The signal emerging from the illuminated focal-spot is collected by a single detector through a confocal pinhole, whose small size improves the resolution but reduces the collection-efficiency.[17] Mathematically, confocal imaging results in an effective PSF, $h_{eff}(r)$, that is equal to the product of the illumination and the detection PSFs[25]

$$h_{eff}(r) = h_{ill}(r) \times h_{det}(r), \quad (1)$$

where $r$ is the spatial coordinate, $h_{ill}$ is the illumination PSF, and $h_{det}$ is the detection PSF. The detection PSF size is determined by the detection pinhole size and the diffraction-limit. For simplicity, isoplanatism is assumed in Eq. (1). For identical detection and illumination PSFs having Gaussian profiles, confocal imaging, thus, results in a $\sqrt{2}$ times narrower effective PSF compared to each of these PSFs.

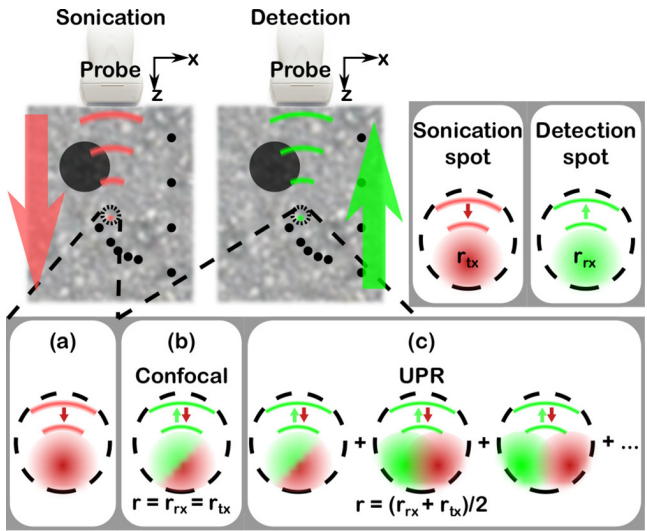

**FIG. 1.** Concept: ultrasound pixel-reassignment (UPR) utilizes the same sonication scheme as in conventional confocal ultrasound imaging (a). In this scheme, the position of a diffraction-limited sonication focus, $r_{tx}$, is scanned over the sample. (b) In confocal detection, only the signals that originate from the sonication position, $r = r_{rx} = r_{tx}$, are used to form the image. (c) UPR utilizes detected signals that originate also outside the sonication position for the reconstruction, improving SNR and resolution. The UPR image at position $r$ is formed by summing all signals that originate from $r_{rx}$ such that $r = (r_{rx} + r_{tx})/2$, i.e., reassigning all the detected signals to the midpoint between sonication and detection positions.

In contrast to confocal microscopy, in ISM, an array of detectors is placed at the image plane instead of the confocal pinhole. The center detector collects the same signal as a conventional confocal detector. However, the neighboring detectors collect the off-focus signals that are usually rejected in confocal detection. To construct the ISM image, at each illumination point, $r_{ill}$, the off-focus signals from each detector position, $r_{det}$, are reassigned to the midpoint between illumination and detection positions:[19,26] $r_{ISM} = \frac{1}{2}(r_{ill} + r_{det})$. The reassigned signals are summed over all scan positions, resulting in an effective PSF that is the convolution of the illumination and detection PSFs, scaled by a factor of half[19,21]

$$h_{ISM}(r) = h_{ill}(2r) \otimes h_{det}(2r), \quad (2)$$

where $\otimes$ denotes convolution.

The main advantage of ISM over confocal imaging is its higher SNR since the photons outside the confocal pinhole are detected and used for imaging instead of being rejected. Additionally, the resolution of ISM is usually better than that of conventional confocal microscopy performed with a pinhole size of one Airy unit.[22,23] The price for the improved resolution and SNR in the optical implementation of ISM is the use of an array of detectors in lieu of a single detector. Since conventional ultrasound imaging is, in fact, a confocal imaging technique performed with an array of detectors,[24] it is natural to apply ISM to ultrasound imaging.

In ultrasound imaging, an array of transducers serves as both the transmitters and the detectors. A reflection image of the medium is formed by transmitting a focused ultrasound wave to each position in the FOV, $r_{tx}$. The reflected waves from each point in the FOV, $r_{rx}$, are recorded by the entire array of transducers. The recorded waveforms from each sonicated position, $r_{tx}$, are digitally delayed-and-summed to reconstruct the time-gated reflected field from each point in the FOV, $r_{rx}$. The entire dataset in such a pulse-echo acquisition is given by the reflection-matrix,[11] $R(r_{tx}, r_{rx})$. The confocal image is formed by detecting the signals that emerge from the sonication point ($r_{rx} = r_{tx}$), i.e., the diagonal elements of the reflection matrix provide the confocal image. In state-of-the-art ultrafast ultrasound systems, the reflection-matrix is built from plane-waves transmissions that are digitally coherently compounded.[11,12,24]

When the sonication field is focused to each point, $r_{tx}$, the spread of the focused field around the point is given by the sonication amplitude point-spread-function (APSF): $E(r, r_{tx})$. Equivalently, in detection, the APSF around the detection point, $r_{rx}$, is given by the detection APSF: $U(r, r_{tx})$. Thus, the reflection-matrix, $R(r_{tx}, r_{rx})$, for a medium with spatial reflectivity, $o(r)$, is given by

$$R(r_{tx}, r_{rx}) = \int o(r') E(r', r_{tx}) U(r', r_{rx}) dr'. \quad (3)$$

A UPR image can be formed by assigning to each imaged point, $r$, all the collected signals that have a sonication point, $r_{tx}$, and a detection point, $r_{rx}$, that $r$ is the midpoint between them, i.e., all the elements of the matrix that fulfill: $r = \frac{1}{2}(r_{rx} + r_{rx})$. Thus, the UPR image is given by

$$I_{UPR}(r) = \int dr_{tx} R(r_{tx}, 2r - r_{tx}). \quad (4)$$

Interestingly, the pixel-reassignment image formation [Eq. (4)] at a given depth, $z$, is a sum over the antidiagonal elements of the reflection matrix, $R(x_{tx}, x_{rx}; z)$.

Considering imaging with a linear transducer-array over a small FOV, at a sufficiently large depth, the APSF of both the sonication, $E(r, r_{tx})$, and detection, $U(r, r_{tx})$, that form the reflection matrix [Eq. (3)], can be considered to be the product of a shift-invariant envelope ($E_{env}$ and $U_{env}$, respectively,) and a nonshift-invariant phase term, $F$ (see theoretical derivation in the supplementary material, Sec. B). Under this approximation, similar to ISM, the UPR image is given by the convolution of the spatial reflectivity, $o(r)$, with an effective APSF that is the scaled convolution: $\hat{E}(2r) \otimes U_{env}(2r)$, where $\hat{E}(r) = E_{env} \times \exp(-ik_0 x^2/z)$ is the sonication APSF envelope multiplied by a parabolic phase term and $U_{env}$ is the detection APSF envelope (see theoretical derivation in the supplementary material, Sec. C)

$$I_{UPR}(r) = o(r) \otimes \left(\hat{E}(2r) \otimes U_{env}(2r)\right). \quad (5)$$

This result is identical to the conventional ISM image in fluorescence microscopy up to the additional phase terms. It can be compared to the conventional confocal ultrasound image, which is the convolution of the spatial reflectivity multiplied by a parabolic phase term, $\hat{o}(r) = o(r) \times \exp(ik_0 x^2/z)$, with the (unscaled) product of the sonication and detection APSFs' envelope

$$I_{conf}(r) = R(r = r_{tx} = r_{rx}) = \hat{o}(r) \otimes (E_{env}(r) \times U_{env}(r)). \quad (6)$$

Considering diffraction-limited APSFs that have a lateral sinc profile, the UPR APSF lateral profile [Eq. (5)] would be the sinc profile scaled by a factor of half, while the confocal APSF lateral profile [Eq. (6)] would be a sinc function squared. Thus, the UPR APSF for this diffraction-limited case is expected to have a smaller FWHM, as we demonstrate experimentally below (Fig. 2). In addition to the gain in resolution, the pixel-reassignment processing is expected to increase the SNR compared to conventional confocal imaging since it utilizes off-focus signals. We experimentally demonstrate the simultaneous SNR increase below [Fig. 3(c)].

To experimentally study the potential of UPR, we have performed a set of imaging experiments on tissue-mimicking acoustic phantoms (GAMMEX SONO404 and SONO403). Data were acquired using a Verasonics P4–2v probe at a center frequency of $f_0$ = 2.7 MHz, having 64 elements with a total aperture size of $D$ = 19.2 mm, connected to a Verasonics Vantage 256 multi-channel system. Sonication was performed by sequential transmission of 65 plane-waves, spanning an angular range of $\theta \in [-18°, 18°]$ with even angular spacing.

Sample results of imaging point reflectors (pins with a diameter of 0.1 mm) at different depths comparing conventional plane wave compounding (confocal ultrasound imaging) to UPR are presented in Fig. 2, along with a comparison to the theoretically expected APSFs [Fig. 2(e)]. In Fig. 2, the UPR image was calculated by summing only the detected waves that originate from a maximum distance of ∼10 mm from the sonication point since the signal to noise at farther away points is low (see analysis of SNR and filter size below). As expected, UPR shrinks the APSF mainlobe transverse width by a factor of approximately 25% at depths larger than 20 mm [Figs. 2(a)–2(c)]. This comes at the price of increased sidelobes [Figs. 2(d) and 2(e)], as expected from the theoretical analysis assuming a sinc-shaped APSF. At shallower depths, the improvement is less significant since the APSF lateral profile differs from a sinc function (see the supplementary material, Sec. B).

As mentioned above, the UPR resolution and SNR are affected by the maximum distance from the sonication point that is used for the pixel-reassignment calculation. We define this maximum distance as the UPR filter size. The filter size is analogous to the size of the focal-plane array used in conventional optical ISM.[26] To analyze the effect of UPR filter size on the resolution and SNR, we have plotted the −10 dB lateral width of the imaged point reflectors and the SNR as a function of the filter size. The results of this investigation are displayed in Fig. 3. The filter size is measured by the maximum distance between sonication and detection points (the filter diameter, $\sigma_{filter}$). In Fig. 3, we plot the resolution and SNR as a function of $\sigma_{filter}$ normalized by the lateral width of a diffraction-limited APSF ($\sigma_x$)

$$\sigma_x = 2\frac{c_0 z}{f_0 D}, \quad (7)$$

where $c_0$ is the speed of sound in the medium, $f_0$ is the center frequency of transmission, $D$ is the probe's width, and $z$ is the examined depth.

As can be seen in Fig. 3(a), the maximal improvement in lateral resolution is obtained for filter sizes larger than ∼3 times the diffraction-limited APSF width, providing a ∼25% improvement in the −10 dB lateral width. The axial resolution shows a small improvement, which is not obtained in optical ISM and is a result of the time-gated ultrasound detection.

To investigate the improvement in SNR as a function of the filter size, we acquired 20 consecutive frames (at 6 fps) of the tissue-mimicking phantom. SNR was calculated as the ratio between the temporal average of the reconstructed signal (square of the envelope absolute value) divided by the temporal standard-deviation of the signal, for each pixel in an imaged patch that does not contain point reflectors. Figure 3(c) presents the average SNR over such 15 × 15 mm² patch at depths of 83 and 140 mm. The optimal improvement in SNR is obtained for a filter size of ∼1 − 1.5 times the diffraction-limited APSF width, providing a ∼1.5 dB and a ∼3 dB improvement, respectively. As expected, for too small filter size, the SNR improvement is lower since less signal is accumulated, whereas for too large filter size the SNR is also lowered, as lower signal energy is collected at points that are far from the sonication point. We attribute the differences between the SNR improvement at different depths and the anisotropy of the APSFs, which grows at larger depths [Fig. 2(a)], whereas the UPR filter used is isotropic.

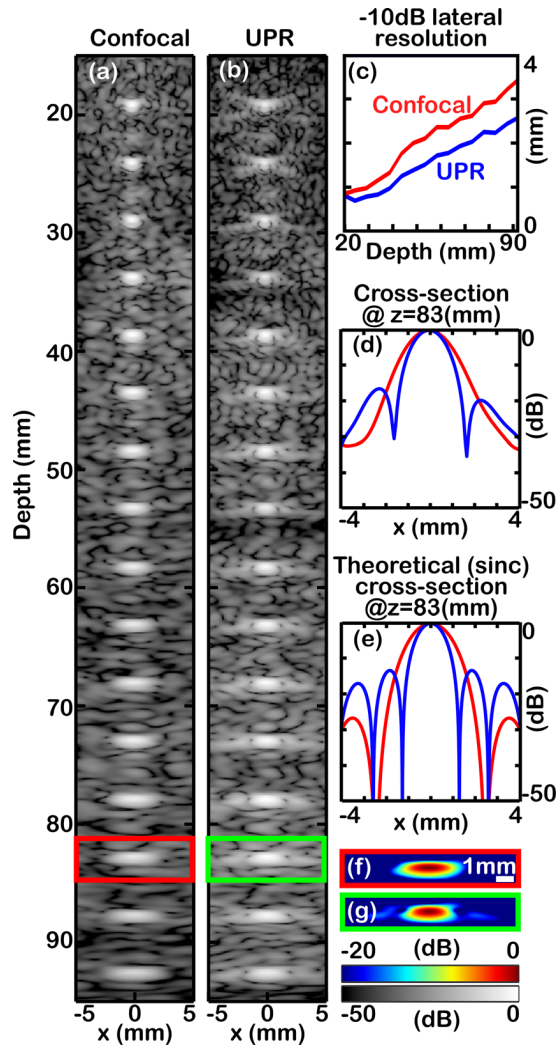

FIG. 2. Experimental comparison of conventional confocal imaging and UPR imaging of a tissue-mimicking acoustic phantom: (a) confocal image. (b) UPR image. (c) Lateral −10 dB width (resolution) of the imaged point reflectors at different depths, for confocal imaging (red), and UPR imaging (blue). (d) A cross section of the point-reflector at a depth of $z$ = 83 mm. (e) Theoretical cross section of a point-reflector at a depth of $z$ = 83 mm, assuming diffraction-limited sinc-shaped APSFs. (f) and (g) Zoom-in on the point-reflector image at a depth of 83 mm in conventional confocal imaging and UPR imaging, respectively. The region-of-interest is marked in (a) and (b) by red and green rectangles, respectively. Data were acquired using a probe with a center frequency $f_0$ = 2.7 MHz and an aperture size of $D$ = 19.2 mm. UPR included signals detected at a maximum distance of ∼10 mm from the sonication point.

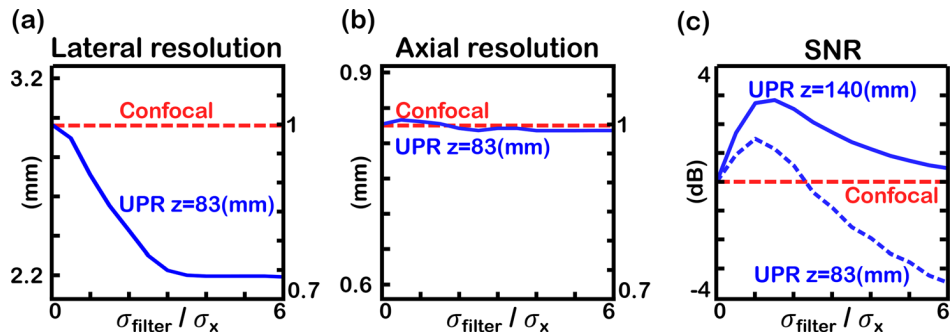

FIG. 3. Experimental characterization of the dependence of the obtained resolution and SNR on the pixel-reassignment filter size, $\sigma_{filter}$. (a) Lateral resolution ($-10$ dB width of a point reflector image) at a depth of $z = 83$ mm, for UPR (blue) and confocal imaging (dashed red). Filter size, $\sigma_{filter}$ is normalized by the lateral width of the theoretical diffraction-limited APSF, $\sigma_x$. (b) Same as (a) for axial resolution. (c) SNR as a function of filter size for UPR at a depth of $z = 83$ mm (dashed blue line), and $z = 140$ mm (solid blue), compared to SNR of confocal imaging (dashed red). Data were acquired as in Fig. 2.

To experimentally demonstrate UPR imaging in different imaging scenarios, we have performed an additional set of imaging experiments on tissue-mimicking acoustic phantoms. Data were acquired as detailed above using a Verasonics P4–2v probe at a center frequency of $f_0 = 2.7$ MHz, having 64 elements with a total aperture size of $D = 19.2$ mm. This probe was used to image a large gray-scale target at a depth of 60 mm (Fig. 4), whose reflectivity is larger than the speckle background by $\sim$6 dB. The second imaging scenario was of a resolution target group of five pins with a diameter of 0.1 mm (Fig. 5). The imaged FOV was at a depth of 60 mm, laterally centered 3.5 mm right from the middle of the probe. The ground truth position of the pins is presented by red "X"s. These results demonstrate an improvement in resolution in more complex scenarios, as can be seen from the decrease in the speckle grain size (Fig. 4) and the improved imaging of the point-reflectors (Fig. 5). However, in high clutter scenes such as Fig. 4, the increased side-lobes may reduce the contrast-to-background ratio. This may not be a significant disadvantage in low-clutter scenarios such as in ultrasound localization microscopy[27] (as discussed below).

In conclusion, performing pixel-reassignment on the coherent fields detected in ultrasound imaging allows a resolution and SNR increase using the conventionally acquired signals. However, achieving optimal resolution improvement without sacrificing SNR requires a careful choice of the UPR filter width (Fig. 3). In our proof-of-principle demonstrations, we have used an isotropic UPR filter; however, an anisotropic UPR filter that is matched to the anisotropic APSF at each depth may result in a further improvement of the SNR. Our proof-of-principle demonstrations were obtained on simple tissue-mimicking phantoms, where the gain in using UPR is most notable when imaging bright point-like reflectors, and the increased sidelobes of UPR have a small effect on the imaging quality (e.g., Fig. 5). This may be especially attractive for ultrasound localization microscopy (ULM), where relatively sparse point-like reflectors are localized.[27] In such scenarios, the width of the APSF mainlobe, which is reduced in UPR, may be more significant for the localization accuracy than the spread of the energy in the APSF sidelobes. UPR may, thus, hold a potential improvement for ULM, assuming a sufficient number of plane-waves sonications is possible. The potential impact on in vivo imaging is left for future work.

Similar to other matrix-based approaches [e.g., spatiotemporal matrix image formation (SMIF)[7]] UPR provides improved resolution and SNR by off-line data processing, at the price of increased computation time. However, unlike SMIF, UPR does not require or uses any prior knowledge of the imaging APSF or object features. The SNR and resolution improvement in UPR do not originate from a more accurate physical model but due to the use of the nonconfocal signals. Techniques, such as SMIF, may have a potential advantage over UPR if the imaging system impulse response can be accurately simulated and priors on the imaged scene are known, at the price of an additional computational burden. The increased computation time compared to confocal image reconstruction is the result of calculating the off diagonal elements in the reflection matrix, which are not required for confocal imaging. The required improvement in computation time is left for future work.

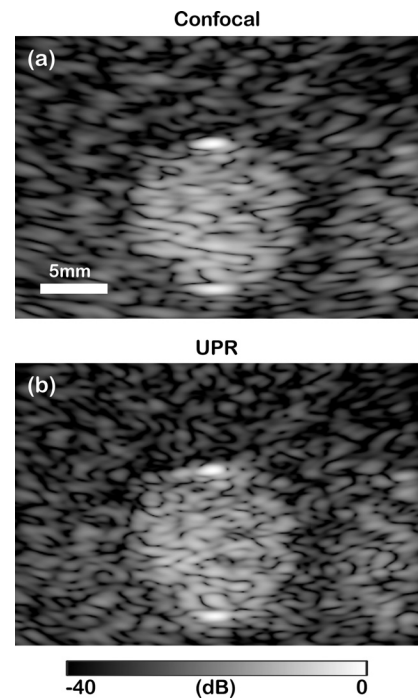

FIG. 4. Experimental comparison of confocal (a) and UPR imaging (b) of a large gray-scale target located at a depth of 60 mm in a tissue-mimicking acoustic phantom. The smaller speckle grain in UPR can be observed. Data were acquired as in Fig. 2.

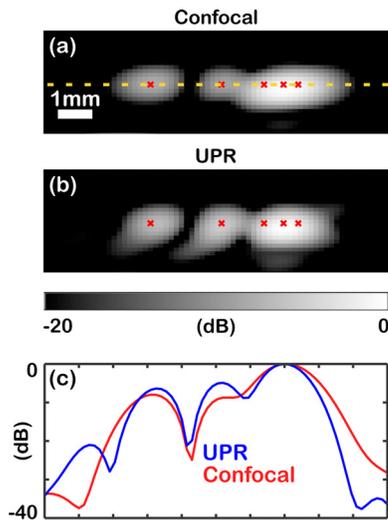

**FIG. 5.** Experimental comparison of confocal (a) and UPR imaging (b) of a resolution target group of five pins with a diameter of 0.1 mm, at a depth of 60 mm in a tissue-mimicking acoustic phantom. Data were acquired as in Fig. 2. The frame center is laterally shifted off-axis by 3.5 mm. The ground truth position of the pins is marked by red Xs. A cross section is presented in (c) and is marked by a dashed yellow line in (a).

See the supplementary material for the theoretical derivations of the coherent sonication and reconstruction APSFs and of the image formation in UPR.

This work has received funding from the Israel Science Foundation (Grant No. 1361/18), the European Research Council (ERC) Horizon 2020 research and innovation program (No. 677909), and was supported by the Ministry of Science and Technology in Israel.

## DATA AVAILABILITY

The data that support the findings of this study are available from the corresponding author upon reasonable request.

## REFERENCES

[1]N. Jiménez, J. Redondo, V. Sánchez-Morcillo, F. Camarena, Y. Hou, and E. E. Konofagou, "Nonlinear acoustics fdtd method including frequency power law attenuation for soft tissue modeling," arXiv:1401.6669 (2014).
[2]A. Fenster and J. C. Lacefield, *Ultrasound Imaging and Therapy* (Taylor & Francis, 2015).
[3]T. Taxt and J. Strand, "Two-dimensional noise-robust blind deconvolution of ultrasound images," IEEE Trans. Ultrason., Ferroelectr., Freq. Control **48**, 861–866 (2001).
[4]G. Matrone, A. S. Savoia, G. Caliano, and G. Magenes, "The delay multiply and sum beamforming algorithm in ultrasound B-mode medical imaging," IEEE Trans. Med. Imaging **34**, 940–949 (2015).
[5]F. Vignon and M. R. Burcher, "Capon beamforming in medical ultrasound imaging with focused beams," IEEE Trans. Ultrason., Ferroelectr., Freq. Control **55**, 619–628 (2008).
[6]N. Wagner, Y. C. Eldar, and Z. Friedman, "Compressed beamforming in ultrasound imaging," IEEE Trans. Signal Process. **60**, 4643–4657 (2012).
[7]B. Berthon, P. Morichau-Beauchant, J. Porée, A. Garofalakis, B. Tavitian, M. Tanter, and J. Provost, "Spatiotemporal matrix image formation for programmable ultrasound scanners," Phys. Med. Biol. **63**, 03NT03 (2018).
[8]T. Ilovitsh, A. Ilovitsh, J. Foiret, B. Z. Fite, and K. W. Ferrara, "Acoustical structured illumination for super-resolution ultrasound imaging," Commun. Biol. **1**, 1–11 (2018).
[9]P. Kruizinga, P. van der Meulen, A. Fedjajevs, F. Mastik, G. Springeling, N. de Jong, J. G. Bosch, and G. Leus, "Compressive 3D ultrasound imaging using a single sensor," Sci. Adv. **3**, e1701423 (2017).
[10]J. Janjic, P. Kruizinga, P. Van Der Meulen, G. Springeling, F. Mastik, G. Leus, J. G. Bosch, A. F. van der Steen, and G. van Soest, "Structured ultrasound microscopy," Appl. Phys. Lett. **112**, 251901 (2018).
[11]W. Lambert, L. A. Cobus, T. Frappart, M. Fink, and A. Aubry, "Distortion matrix approach for ultrasound imaging of random scattering media," Proc. Natl. Acad. Sci. **117**, 14645 (2020).
[12]W. Lambert, L. A. Cobus, M. Couade, M. Fink, and A. Aubry, "Reflection matrix approach for quantitative imaging of scattering media," Phys. Rev. X **10**, 021048 (2020).
[13]H. Bendjador, T. Deffieux, and M. Tanter, "The SVD beamformer: Physical principles and application to ultrafast adaptive ultrasound," IEEE Trans. Med. Imaging **39**, 3100–3112 (2020).
[14]M. Jaeger, E. Robinson, H. G. Akarçay, and M. Frenz, "Full correction for spatially distributed speed-of-sound in echo ultrasound based on measuring aberration delays via transmit beam steering," Phys. Med. Biol. **60**, 4497 (2015).
[15]R. Rau, D. Schweizer, V. Vishnevskiy, and O. Goksel, "Ultrasound aberration correction based on local speed-of-sound map estimation," in *2019 IEEE International Ultrasonics Symposium (IUS)* (IEEE, 2019), pp. 2003–2006.
[16]J. Shepherd, G. Renaud, P. Clouzet, and K. van Wijk, "Photoacoustic imaging through a cortical bone replica with anisotropic elasticity," Appl. Phys. Lett. **116**, 243704 (2020).
[17]J. G. Fujimoto and D. Farkas, *Biomedical Optical Imaging* (Oxford University Press, 2009).
[18]J. Mertz, *Introduction to Optical Microscopy* (Cambridge University Press, 2019).
[19]C. B. Müller and J. Enderlein, "Image scanning microscopy," Phys. Rev. Lett. **104**, 198101 (2010).
[20]R. Tenne, U. Rossman, B. Rephael, Y. Israel, A. Krupinski-Ptaszek, R. Lapkiewicz, Y. Silberberg, and D. Oron, "Super-resolution enhancement by quantum image scanning microscopy," Nat. Photonics **13**, 116–122 (2019).
[21]C. R. Sheppard, "Super-resolution in confocal imaging," Optik (Stuttgart) **80**, 53–54 (1988).
[22]C. J. Sheppard, S. B. Mehta, and R. Heintzmann, "Superresolution by image scanning microscopy using pixel reassignment," Opt. Lett. **38**, 2889–2892 (2013).
[23]S. Roth, C. J. Sheppard, K. Wicker, and R. Heintzmann, "Optical photon reassignment microscopy (OPRA)," Opt. Nanosc. **2**, 1–6 (2013).
[24]G. Montaldo, M. Tanter, J. Bercoff, N. Benech, and M. Fink, "Coherent plane-wave compounding for very high frame rate ultrasonography and transient elastography," IEEE Trans. Ultrason., Ferroelectr., Freq. Control **56**, 489–506 (2009).
[25]C. Sheppard and A. Choudhury, "Image formation in the scanning microscope," Opt. Acta **24**, 1051–1073 (1977).
[26]C. J. Sheppard, M. Castello, G. Tortarolo, T. Deguchi, S. V. Koho, G. Vicidomini, and A. Diaspro, "Pixel reassignment in image scanning microscopy: A re-evaluation," J. Opt. Soc. Am. A **37**, 154–162 (2020).
[27]C. Errico, J. Pierre, S. Pezet, Y. Desailly, Z. Lenkei, O. Couture, and M. Tanter, "Ultrafast ultrasound localization microscopy for deep super-resolution vascular imaging," Nature **527**, 499–502 (2015).